\documentclass[12pt]{article}

\usepackage{amsmath,amssymb,array,amsfonts}
\usepackage[dvipdfmx]{graphicx}
\usepackage{comment,color}
\usepackage{cite}

\newcommand{\bs}{\boldsymbol}

\allowdisplaybreaks

\def\ncm{\newcommand}
\def\e {{\rm e}}
\def\s {{\rm s}}
\def\H {{\rm H}}
\def\SM{{\rm SM}}

\def\Tr{{\rm Tr}}

\def\bls{\baselineskip}
\def\dis{\displaystyle}
\def\lra{\longrightarrow}

\def\dy{\int_{-\pi R}^{\pi R}\!\!\!\!\!\!\!\!{\rm d}y\,}%

\ncm{\sls}[1]{{\ooalign{\hfil/\hfil\crcr$#1$}}}

\makeatletter
    
    \@addtoreset{equation}{section}
\makeatother

\begin{document}
\setlength{\baselineskip}{18pt}
\begin{titlepage}

\begin{flushright}
KOBE-TH-11-08
\end{flushright}
\vspace{1.0cm}
\begin{center}
{\Large\bf $\bs{B^0}$\,--\,$\bs{\bar B^0}$ Mixing in Gauge-Higgs Unification} 
\end{center}
\vspace{25mm}

\begin{center}
{\large
Yuki Adachi, 
%\footnote{e-mail : yuki1983@kobe-u.ac.jp},
%
Nobuaki Kurahashi$^*$, 
%\footnote{e-mail : 075s112s@stu.kobe-u.ac.jp},
%
Nobuhito Maru$^{**}$
%\footnote{e-mail : lim@kobe-u.ac.jp}
%
and Kazuya Tanabe$^*$
}
\end{center}
\vspace{1cm}
\centerline{{\it
Department of Sciences, Matsue College of Technology,
Matsue 690-8518, Japan.}}

\centerline{{\it
$^*$Department of Physics, Kobe University,
Kobe 657-8501, Japan.}}

\centerline{{\it
$^{**}$Department of Physics, and Research and Education Center for Natural Sciences,}}
\centerline{{\it
Keio University,
Yokohama 223-8521, Japan.
}}
%
%%%%%%%%%%%%%%%%%%%%%%%%%%%%%%% Abstract %%%%%%%%%%%%%%%%%%%%%%%%%%%%%%%
%
\vspace{2cm}
\centerline{\large\bf Abstract}
\vspace{0.5cm}

We discuss flavor mixing and resulting Flavor Changing Neutral Current (FCNC) 
in a five dimensional $SU(3)_{\rm color} \otimes SU(3)\otimes U'(1)$ gauge-Higgs unification.
Flavor mixing is realized by the fact
that the bulk and brane localized mass terms are not diagonalized simultaneously. 
%unless bulk masses are degenerate.
%Thus the FCNC process disappears for degenerate bulk masses.
As the concrete FCNC processes, 
we calculate the rate of $B_d^0$\,--\,$\bar B_d^0$ mixing and $B_s^0$\,--\,$\bar B_s^0$ mixing
due to the exchange of non-zero Kaluza-Klein gluons at the tree level. 
We obtain a lower bound on the compactification scale of order $\cal O$(TeV)
by comparing our prediction on the mass difference of neutral $B$ meson
with the recent experimental data. 
%which is much milder than what we naively expect assuming only the decoupling
%of non-zero Kaluza-Klein gluons. 

\end{titlepage}

%\maketitle
%\thispagestyle{empty}

%\vspace{5mm}

%}

\newpage
%%% Introduction %%%%%%%%%%%%%%%%%%%%%%%%%%%%%%%%%%%%%%%%%%%%%%%%%%%%%%%%%%
\section{Introduction}
In spite of the great success of the Standard Model (SM),
the origin of electroweak gauge symmetry breaking is still unknown in particle physics.
Though in the SM, Higgs boson is assumed to play a role for the symmetry breaking,
it seems to have various theoretical problems such as the hierarchy problem
and the presence of many theoretically unpredicted arbitrary coupling constants
in its interactions.

Gauge-Higgs unification (GHU) \cite{GH}
is one of the fascinating scenarios beyond the SM.
It provides a possible solution to the hierarchy problem without supersymmetry. 
%and reduces the number of arbitrary couplings in Higgs sector. 
%also shedding some light on the long standing arbitrariness problem of Higgs interactions.
In this scenario, Higgs boson in the SM is identified
with the extra spatial components of the higher dimensional gauge fields.
A remarkable fact is that the quantum correction to Higgs mass is UV-finite
and calculable due to the higher dimensional gauge symmetry 
regardless of the non-renormalizability of the theory.
%This feature is guaranteed by the higher dimensional gauge invariance and 
This has opened up a new avenue to solve the hierarchy problem \cite{HIL}.
The finiteness of the Higgs mass has been studied and verified in various models
and types of compactification at one-loop level%
\footnote{For the case of gravity-gauge-Higgs unification, see \cite{HLM}.} \cite{ABQ}
and even at the two loop level \cite{MY}.
The fact that the Higgs boson is a part of gauge fields implies that
Higgs interactions are restricted by gauge principle
and may provide a possibility to solve the arbitrariness problem of Higgs interactions as well.

{}From such point of view, 
it seems that the following issues are particularly important for the GHU to be phenomenologically viable. 
The first one is whether there are any characteristic prediction on the observables subject to precision tests. 
The second one is how CP violation is achieved since the Higgs interactions are given by gauge interactions with real couplings. 
The last one is how flavor mixing is generated since Yukawa coupling in GHU is given by gauge interactions which are universal for all flavors. 
%\begin{enumerate}

%\item[(1).]
%Are there any characteristic and generic predictions on the observables,
%which are subject to precision tests\,?

%\item[(2).]
%How are the flavor structure of fermion masses and flavor mixings realized
%in the Yukawa couplings starting from higher dimensional gauge interaction\,?

%\item[(3).]
%In view of the fact that Higgs interactions are basically gauge interactions
%with real gauge coupling constants, how is CP violated\,?

%\end{enumerate}
%\noindent 
%Let us note that the problems (2) and (3) are also shared by superstring theories,
%where the low energy effective theory, 
%i.e. the point particle limit,
%of the open string sector is (10-dimensional) super Yang-Mills theories,
%which can be regarded as a sort of GHU.  
As for the first issue,
it will be desirable to find finite (UV-insensitive) and calculable observables,
in spite of the fact that
the theory is non-renormalizable and observables are very UV-sensitive in general.
Works on the oblique electroweak parameters and fermion anomalous magnetic moment
from such a viewpoint have been already done in the literature \cite{LM, LHC, ALM1}.
The second issue has been addressed in our previous papers \cite{ALM3, LMN},
where CP violation is claimed to be achieved spontaneously
either by the VEV of the Higgs field
or by the complex structure of the compactified extra space.

In this paper,
we focus on the remaining issue concerning the flavor physics in the GHU scenario.
It is highly non-trivial problem to explain the variety of fermion masses 
and flavor mixings in this scenario,
since the gauge interactions should be universal for all matter fields,
while the flavor symmetry has to be broken eventually
in order to distinguish each flavor and to realize their mixings.
In our previous papers \cite{2010AKLM, 2011AKLM},
we addressed this issue and have clarified the mechanism to generate the flavor mixings
by the interplay between bulk masses and the brane localized masses.
%As a remarkable property of higher dimensional gauge theories
%gauge invariant bulk mass terms are allowed
%in the form of sign functions of the extra space coordinate $y$.
%In the SM $SU(2)$ left-handed doublet of fermions couples
%to both of up-type and down-type right-handed fermions simultaneously.
%In the GHU, however, the up- and down-type right-handed fermions
%belong to different representations of gauge group, in general.
%Thus they couple to two independent left-handed doublets to form Yukawa couplings
%and therefore it is needed to introduce brane-localized fermions
%such that they form brane-localized mass terms
%with some linear combinations of the doublets
%in order to eliminate the redundant degrees of freedom.

Important point is that
such introduced two types of mass terms generically may be flavor non-diagonal
without contradicting with gauge invariance, 
which leads to the flavor mixing
in the up- and down-types of Yukawa couplings \cite{BN}.
We may start with the base where the bulk mass terms are diagonalized,
since the bulk mass terms are written in the form of hermitian matrix,
which may be diagonalized by suitable unitary transformations,
keeping the kinetic and gauge interaction terms of fermions invariant
\cite{2010AKLM, 2011AKLM}.
Even in this base, however, the brane-localized mass terms still have off-diagonal elements
in the flavor base in general.
Namely, the fact that two types of fermion mass terms cannot be diagonalized simultaneously
leads to physical flavor mixing.
%This is why we stress that the interplay between these two types of mass terms is crucial.  

%At first thought,
%one might think that only the brane localized mass terms are enough
%to generate the flavor mixings since they can be put by hand.
%However, it is not the case.
%We have shown that the flavor mixings disappear in the limit of universal bulk masses
%where the hierarchy of fermion masses is absent \cite{2010AKLM, 2011AKLM}.
%The reason is in this limit the bulk mass terms remain flavor-diagonal
%for arbitrary unitary transformation of each representation of bulk fermions.
%By use of this degree of freedom the Yukawa couplings are readily made diagonal.
%This is a remarkable feature of the GHU scenario,
%which is not shared by, e.g., the universal extra dimension 
%where the flavor mixing may be caused by Yukawa couplings in the bulk
%just as in the standard model.

Once the flavor mixings are realized,
it will be important to discuss flavor changing neutral current (FCNC) processes,
which have been playing a crucial role
for checking the viability of various new physics models,
as is seen in the case of SUSY model.
This issue was first discussed in \cite{DPQ} in the context of extra dimensions.
Since our model reduces to the SM at low energies,
there is no FCNC processes at the tree level with respect to the zero mode fields.
However, it turns out that
the exchange of non-zero Kaluza-Klein (KK) modes of gauge bosons causes FCNC
at the tree level,
though the rates of FCNC are suppressed
by the inverse powers of the compactification scale (\lq\lq decoupling")
\cite{2010AKLM, 2011AKLM}.
The reason is the following.  
%As a genuine feature of the higher dimensional gauge theories with orbifold compactification,
%the gauge invariant bulk mass terms for fermions,
%generically written as $M\epsilon(y)\bar\psi\psi$ with $\epsilon(y)$
%being the sign function of extra space coordinate $y$, are allowed.
%The bulk mass $M$ may be different depending on each generation
%and can be an important new source of the flavor violation.
%The presence of the mass terms causes the localization of Weyl fermions
%in two different fixed points of the orbifold depending on their chiralities
%and the Yukawa coupling obtained by the overlap integral over $y$
%of the mode functions of Weyl fermions with different chiralities
%is suppressed by a factor $2\pi RM\e^{-\pi RM}$ ($R$ : the size of the  extra space),
%which is otherwise just gauge coupling $g$ and universal for all flavors.
%Thus in GHU scenario, fermion masses are all equal and of weak scale $M_W$ to start with
%and the observed hierarchical small fermion masses can be achieved without fine tuning
%thanks to the exponential suppression factor $\e^{-\pi RM}$.
%On the other hand,
%this means that the criteria by Glashow-Weinberg \cite{GW} in 4D space-time is not enough
%to ensure natural flavor conservation.
%Namely, 
The gauge couplings of non-zero KK modes of gauge boson,
whose mode functions are $y$-\hspace{0mm}dependent,
to zero mode fermions are no longer universal 
%even for Weyl fermions with definite chirality and the same quantum numbers,
since the overlap integral of mode function of fermion and KK gauge boson depends
on the bulk mass $M$ different from flavor by flavor in general.
%Thus once we move to the base of mass-eigenstates FCNC appears at the tree level.

In the previous papers, as typical processes of FCNC,
we have calculated the $K^0$\,--\,$\bar K^0$ mixing and the $D^0$\,--\,$\bar D^0$ mixing amplitude
at the tree level via non-zero KK gluon exchange
and obtained the lower bounds for the compactification scale
as the predictions of our model \cite{2010AKLM, 2011AKLM}.
Interestingly,
the obtained lower bounds of $\mathcal O(10)$\,TeV were much milder
than what we naively expect assuming that the amplitude is simply suppressed
by the inverse powers of the compactification scale, say $\mathcal O(10^3)$\,TeV.
We pointed out the presence of suppression mechanism of the FCNC processes,
which is operative for light fermions in the GHU model.
%As was mentioned above,
%fermion masses much smaller than $M_W$ are realized by the localizations of fermions.
%Larger the bulk mass $M$, the localization of fermion is steeper
%and therefore for the fermions the mode functions of KK gluons seem to be almost constant.
%Thus for light fermions the gauge couplings of KK gluons become almost universal,
%just as in the case of the zero-mode sector.
In the analysis,
we focused on the simplified two generation scheme
in order to estimate the mass difference and the lower bound on the compactification scale. %$R^{-1}$. 

On the other hand, 
these suppression mechanism in the third generation containing top and bottom quarks does not work so strongly  
by the absence of bulk masses as we will discuss in the main text. 
Then it is expected that the dangerous large FCNC containing the third generation such as $B^0-\bar B^0$ mixing arises 
and more stringent constraints will be obtained.
Thus it would be more desirable to discuss the FCNC process in the three generation scheme.

In this paper, we discuss flavor mixings in the three generation model
and especially consider the typical FCNC processes,
i.e. $B_d^0$\,--\,$\bar B_d^0$ mixing and $B_s^0$\,--\,$\bar B_s^0$ mixing,
which is caused by the mixing between down and bottom quarks or strange and bottom quarks.

We will calculate the dominant contribution to the $B_d^0$\,--\,$\bar B_d^0$ mixing
and the $B_s^0$\,--\,$\bar B_s^0$ mixing %the processes 
at the tree level by the non-zero KK gluon exchange. 
The rate of the FCNC processes is suppressed 
by the small mixings between the third generation and lighter generations. 
Comparing the prediction of our model with the data, 
the lower bound on the compactification scale is obtained.

%%% comment %%%%%%%%%%%%%%%%%%%%%%%%%%%%%%%%%%%%%%%%%%%%%%%
\begin{comment}
It will be also discussed how the extent of the suppression of FCNC process is different depending on the type of contributing effective 4-Fermi operators, 
i.e. the operators made by the product of currents with the same chirality (LL and RR type) and different chiralities (LR type).  
\end{comment}
%%%%%%%%%%%%%%%%%%%%%%%%%%%%%%%%%%%%%%%%%%%%%%%%%%%%%%%%%%%

This paper is organized as follows.
After introducing our model in the next section,
we summarize in section 3 how the flavor mixing is realized
in the context of the gauge-Higgs unification,
which was clarified and described in detail in our previous paper \cite{2010AKLM, 2011AKLM}.
In section 4, as an application of the flavor mixing discussed in section 3,
we calculate the mass difference of neutral $B$-\hspace{0pt}mesons
caused by the $B_d^0$\,--\,$\bar B_d^0$ mixing and the $B_s^0$\,--\,$\bar B_s^0$ mixing
via non-zero KK gluon exchange at the tree level.
We also obtain the lower bound for the compactification scale
by comparing the obtained result with the experimental data.
Our conclusion is given in section 5.

%%% comment %%%%%%%%%%%%%%%%%%%%%%%%%%%%%%%%%%%%%%%%%%%%%%%
\begin{comment}
The origin of the suppression mechanism of FCNC process is discussed in section 5,
emphasizing the importance of the localization of quark fields
and the fact that FCNC is controlled by the non-degeneracy of bulk masses,
which is specific to the gauge-Higgs unification.
Also discussed is the origin of the different extent of the suppression depending
on the chirality of the relevant 4-Fermi operator.
\end{comment}
%%%%%%%%%%%%%%%%%%%%%%%%%%%%%%%%%%%%%%%%%%%%%%%%%%%%%%%%%%%

%%% The Model %%%%%%%%%%%%%%%%%%%%%%%%%%%%%%%%%%%%%%%%%%%%%%%%%%%%%%%%%%%%%%%%%%%%%%%%%%%
\section{The Model}
%As was discussed in the introduction,
%we should modify our model in \cite{2010AKLM, 2011AKLM}.
The model we consider in this paper is a five dimensional (5D)
$SU(3)_{\rm color} \otimes SU(3) \otimes U'(1)$ GHU model 
compactified on an orbifold $S^1\!/Z_2$ with a radius $R$ of $S^1$. 
The three generation model is basically obtained by extending our previous model, 
but top quark mass cannot be incorporating as it stands. 
%a serious issue is how to implement the top quark mass.
It is known that the fermion masses have an upper bound in GHU, 
%%% \label{rankfactor} %%%%%%%%%%%%%%%%%%%%%%
\begin{align}
	\label{rankfactor}
	m_q \le \sqrt nM_W
	\qquad\Big(\,M_W\text{ : $W$-boson mass}\,\Big)
\end{align}
%%%%%%%%%%%%%%%%%%%%%%%%%%%%%%%%%%%%%%%%%%%%%%%
where $n$ is the number of indices of the representation the fermion belongs to \cite{2005MSSS}. 
Up-type quarks in our model belong to the totally symmetric tensor representation of $SU(3)$,
i.e. $n=2$, in our two generation model.
Thus, we should modify our model to obtain the correct top mass $m_t \sim 2M_W$.
Obviously, the simplest choice would be a 4-rank tensor representation. 
The representation of rank 4 of $SU(3)$ are known to be 
%with their decomposition under $SU(2)\times U(1)$, are
$\overline{\bs{15}}$, $\overline{\bs{24}}$ and $\bs{27}$ \cite{2006CCP}.
We modify our model by using the smallest representation $\overline{\bs{15}}$.
Although it is still remaining a small gap between top and twice of $W$-boson masses, %$m_t-2M_W$ 
it is attributed to the quantum correction of top Yukawa coupling.
%The $SU(3) \otimes U'(1)$ unifies the electro-weak interactions $SU(2) \otimes U(1)$.
Focusing on the quark sector, 
we introduce three generations of bulk fermion in the $\bs3$, two generations of them in the $\bar{\bs6}$
and one generation of bulk fermion in the $\overline{\bf15}$ dimensional representations
of $SU(3)$ gauge group,
%denoted by a column vector, a $3 \times 3$ matrix and a 4-rank tensor.
\begin{subequations}
\begin{alignat}{3}
	\psi^i(\bs3)
 &= Q_3^i \oplus d^i && \big(\,i = 1, 2, 3\,\big), \\
	\psi^i(\bar{\bs6})
 &= \Sigma_6^i \oplus Q_6^i \oplus u^i && \big(\,i = 1, 2\,\big), \\
	\psi(\overline{\bf15})
 &= \Theta \oplus \Delta \oplus \Sigma_{15} \oplus Q_{15} \oplus t
	 &\qquad& %\big(\,i = 3\,\big)
\end{alignat}
\end{subequations}
where all of the fermions are decomposed into those in the representations of $SU(2)$ subgroup of $SU(3)$ gauge group. 
%we have added a charge under 
An extra $U'(1)$ is required for $\psi(\overline{\bf15})$ to fix the hypercharges.
%and where index $i$ stands for generation.
These sets of fermions contain ordinary quarks of the SM in the zero mode sector,
i.e. 
%a pair of $SU(2)$ doublet, 
$Q_3^i$ and $Q_6^i$ ($i = 1, 2$) corresponding to the first two generation quark doublets, 
$Q_3^{i=3}$ and $Q_{15}$ corresponding to the third generation quark doublet,
and $d^i$ ($i = 1, 2, 3$), $u^i$  ($i = 1, 2$), $t$ corresponding to three generation down-type quark singlets, 
the first two generation up-type quark singlets, top quark singlet, respectively.
$\psi^i(\bar{\bs6})$ have $SU(2)$ triplet exotic states $\Sigma_6^i$
and $\psi~(\overline{\bf15})$ also does
$SU(2)$ quintet, quartet, triplet exotic states $\Theta$, $\Delta$ and $\Sigma_{15}$.

The bulk Lagrangian is given by 
\begin{align}
	\mathcal{L}
 =&	-\!\frac12\Tr\big(F_{MN}F^{MN}\big)-\frac14B_{MN}B^{MN}
	-\frac12\Tr\big(G_{MN}G^{MN}\big)\notag\\*
  &	+\bar\psi^i({\bs3})\big\{i\sls D_3-M^i\epsilon(y)\big\}\psi^i(\bs3)
       +\bar\psi^{i=3}({\bs3}) i\sls D_3 \psi^{i=3}(\bs3) \notag \\
  &	+\bar\psi^i(\bar{\bs6})\big\{i\sls D_6-M^i\epsilon(y)\big\}\psi^i(\bar{\bs6})
  %\notag\\*
  %& 
  +\bar\psi(\overline{\bf15})
	 %\big\{
	 i\sls D'_{15}
	 %-M^3\epsilon(y)\big\}
	 \psi(\overline{\bf15})
	 \qquad
\end{align}
where the gauge kinetic terms for $SU(3), U'(1), SU(3)_{{\rm color}}$ 
 and the covariant derivatives are 
%$i = 1, 2$ and where
\begin{subequations}
\begin{align}
	F_{MN}
 &=	\partial_MA_N-\partial_NA_M-ig\big[A_M,A_N\big]\ ,\\
	B_{MN}
 &=	\partial_MB_N-\partial_NB_M\ ,\\
	G_{MN}
 &=	\partial_MG_N-\partial_NG_M-ig_\s\big[G_M,G_N\big]\ ,\\
	\sls D
 &=	\varGamma^M(\partial_M-igA_M-ig_\s G_M)\ ,\\
	\sls D'
 &=	\varGamma^M(\partial_M-igA_M-ig'B_M-ig_\s G_M)\ .
%	\sls D_3\psi^i(\bs3) 
% &= \varGamma^M(\partial_M-igA_M-ig_\s G_M)\psi^i(\bs3)\ ,\\
%	\sls D_6\psi^i(\bar{\bs6}) 
% &= \varGamma^M
%	\Big[
%	\partial_M\psi^i(\bar{\bs6})
%	+ig\big\{A_M^*\psi^i(\bar{\bs6})+\psi^i(\bar{\bs6})(A_M)^\dagger\big\}
%	-ig_\s G_M\psi^i(\bar{\bs6})
%	\Big]
\end{align}
\end{subequations}
%with $G_M$ being understood to act on the color index, not explicitly written here.
The gauge fields $A_M$ and $G_M$ are written in a matrix form,
e.g.~$A_M = A_M^a \frac{\lambda^a}2$ in terms of Gell-Mann matrices $\lambda^a$.
It should be understood that $A_M$
in the covariant derivative $D_M = \partial_M-igA_M-ig_\s G_M$
acts properly depending on the representations of the fermions.
$M, N = 0, 1, 2, 3, 5$ denotes indices of the bulk space-time. 
The five dimensional gamma matrices are given by $\varGamma^M=(\gamma^\mu, i\gamma^5)$ ($\mu=0, 1, 2, 3$).
$g$, $g'$ and $g_\s$ are 5D gauge coupling constants
of $SU(3)$, $U'(1)$ and $SU(3)_{\rm color}$, respectively.
$M^i$ ($i = 1,2$) are generation dependent bulk mass parameters of the first two generations of fermion
accompanied by the sign function $\epsilon(y)$. 
For the third generation, the bulk mass parameter should be taken to be zero to reproduce top quark mass. 
%As was discussed in the introduction,
%here we take the base where the bulk mass term is flavor-diagonal.

The periodic boundary condition is imposed along $S^1$ and $Z_2$ parity
assignments are taken for gauge fields as
\begin{subequations}
\begin{align}
&A_\mu(-y) = P A_\mu(y) P^{-1}, \quad A_y(-y) = - P A_y(y) P^{-1}, \\
&G_\mu(-y) = G_\mu(y), \hspace*{1.5cm} G_y(-y) = - G_y(y), \\ 
&B_\mu(-y) = B_\mu(y), \hspace*{1.5cm} B_y(-y) = - B_y(y)
\end{align}
\end{subequations}
where the orbifolding matrix is defined as $P={\rm diag}(-,-,+)$ and operated in the same way at the fixed points $y=0, \pi R$. 
We can see that the gauge symmetry $SU(3)$ is explicitly broken
to $SU(2) \times U(1)$ by the boundary conditions.
The gauge fields with $Z_2$ odd parity and even parity are expanded
by use of mode functions,
%which are just trigonometric functions, i.e.
\begin{align}
	S_n(y)
 &= \frac1{\sqrt{\pi R}}\sin\frac nRy
	\quad , \qquad
	C_n(y)
  = 
  %\left\{
	%\begin{array}{lc}
%	 \dis\frac1{\sqrt{2\pi R}} & (n = 0)\\[10pt]
	 %\dis
	 \frac1{\sqrt{2^{\delta_{n,0}} \pi R}}\cos\frac nRy 
	 %& (n \ne 0)
%\end{array}
	%\right.
	,
\end{align}
respectively.

The $Z_2$ parities of fermions are assigned
for each component of the representations as follows:
%\begin{subequations}
\begin{alignat}{3}
	\Psi^i({\bs3})
 =&~\big\{Q_{3L}^i(+,+)+Q_{3R}^i(-,-)\big\}
	\oplus\big\{d_L^i(-,-)+d_R^i(+,+)\big\}
  &\quad& \big(\,i = 1,2,3\,\big)\ ,\notag\\
	\Psi^i(\bar{\bs6})
 =&~\big\{\Sigma^i_{6L}(-,-)+\Sigma^i_{6R}(+,+)\big\}
	\oplus\big\{Q_{6L}^i(+,+)+Q_{6R}^i(-,-)\big\}\notag\\*
  &~\oplus\big\{u^i_L(-,-)+u^i_R(+,+)\big\}
  &\quad& \big(\,i = 1,2\,\big)\ ,\notag\\
	\Psi(\overline{\bf15})
 =&~\big\{\Theta_L(-,-)+\Theta_R(+,+)\big\}
	\oplus\big\{\Delta_L(+,+)+\Delta_R(-,-)\big\}\notag\\*
  &~\oplus\big\{\Sigma_{15L}(-,-)+\Sigma_{15R}(+,+)\big\}
	\oplus\big\{Q_{15L}(+,+)+Q_{15R}(-,-)\big\}\notag\\*
  &~\oplus\big\{t_L(-,-)+t_R(+,+)\big\}. 
  &\quad& %\big(\,i = 3\,\big)
  \notag
\end{alignat}
%\end{subequations}
Thus a chiral theory is realized in the zero mode sector by $Z_2$ orbifolding.

The fermions are also expanded by an orthonormal set of mode functions.
Here we will focus on the zero-mode sector necessary for the argument of flavor mixing. 
%\begin{subequations}
%\begin{align}
%	\psi^i(\bs3)
% &= \left[
%	\begin{array}{c}
%	 Q_{3L}^if_L^i(y) \\[6pt]
%	 d_R^if_R^i(y)
%	 \\
%	\end{array}
%	\right]\ , \\
%%% \label{psi6} %%%%%%%%%%%%%%%%%%%%%%
%	\label{psi6}
%	\psi^i\big(\bar {\bf 6}\big)
% &= \left[
%	\begin{array}{c|c}
%		i\sigma^2\Sigma^i\big(i\sigma^2\big)^{\!\T}
%	 &  \dis\frac1{\sqrt2}i\sigma^2Q_6^i\rule[-14pt]{0pt}{0pt}\\[6pt] \hline
%		\dis\frac1{\sqrt2}(Q_6^i)^{\T}\big(i\sigma^2\big)^{\!\T}\rule[20pt]{0pt}{0pt}
%	 &  u^i
%	\end{array}
%	\right]\ ,\\
%%
%	\psi^i\big(\overline{\bf15}\big)
% &= 
%\end{align}
%\end{subequations}
%where $i\sigma^2$ denotes an $SU(2)$ invariant anti-symmetric tensor
%$\big(i\sigma^2\big)^{\alpha\beta}\! = \epsilon^{\alpha\beta}$.
The zero mode sector of each component
of $\psi^i({\bs3}), \psi^i(\bar{\bs6})$ and $\psi(\overline{\bf15})$ are written
in the following way. 
%in terms of the same mode functions as in the case of $\psi^i(\bs3)$.
\begin{subequations}
\begin{align}
&Q_3^i = Q_{3L}^i f_L^i(y)\ , \qquad d^i=d_R^i f_R^i(y) \qquad \big(\, i=1,2,3\, \big), \\ 
	&\Sigma_6^i
 = \Sigma_{6R}^if_R^i(y)\ ,\qquad
	Q_6^i
  = Q_{6L}^if_L^i(y)\ ,\qquad
	u^i
  = u_{R}^if_R^i(y)
	\qquad\big(\,i = 1,2\,\big) \\
%\end{align}
%and for $i = 3$,
%\begin{gather}
	&\Theta
  = \Theta_R f_R(y)\ , \quad
	\Delta
  = \Delta_L f_L(y)\ , \quad
	\Sigma_{15}
  = \Sigma_{15R} f_R(y)\ , \quad
  %\notag\\[2pt]
	Q_{15}
  = Q_{15L} f_L(y)\ ,\quad
	t
  = t_R f_R(y). 
\end{align}
%\end{gather}
\end{subequations}
The mode function for the zero mode of each chirality is given in \cite{ALM3}:
\begin{align}
	f^i_L(y)
 &= \sqrt{\frac{M^i}{1-\e ^{-2\pi RM^i}}}\e^{-M^i|y|},\quad
	f^i_R(y)
  = \sqrt{\frac{M^i}{\e^{2\pi RM^i}-1}}\e^{M^i|y|}. 
\end{align}

We notice that there are two left-handed quark doublets
$Q_{3L}$ and $Q_{6L}$($Q_{15L}$) per generation in the zero mode sector,
which are massless before electro-weak symmetry breaking.
In the one generation case, for instance,
one of two independent linear combinations of these doublets should correspond
to the quark doublet in the SM,
but the other one should be regarded as an exotic state.
Moreover, having an exotic fermion
$\Sigma_{6R}$, $\Sigma_{15R}$, $\Delta_L$ and $\Theta_R$, 
we therefore introduce brane localized four dimensional Weyl spinors 
to form $SU(2) \times U(1)$ invariant brane localized Dirac mass terms
in order to remove these exotic massless fermions
from the low-energy effective theory \cite{BN, ACP}.
\begin{align}
	\mathcal L_{\rm BLM}
 &= \mathcal L_{\rm BLM}^Q
	+\mathcal L_\text{BLM}^{\Sigma_6}
	+\mathcal L_{\rm BLM}^{\Sigma_{15}}
	+\mathcal L_{\rm BLM}^\Delta
	+\mathcal L_{\rm BLM}^\Theta
\end{align}
where for the first two generations,
\begin{subequations}
\begin{align}
	\mathcal L_{\rm BLM}^{\Sigma_6}
 &= \int^{\pi R}_{-\pi R}\hspace{-5.5mm}dy\,\sqrt{2\pi R}\,
	m_{\rm BLM}^{\Sigma_6}\delta(y-\pi R)\bar\Sigma^i_{6R}(x,y)\Sigma^i_{6L}(x)
	+({\rm h.c.})
\intertext{and for the third generation,
}
	\mathcal L_{\rm BLM}^{\Sigma_{15}}
 &= \int^{\pi R}_{-\pi R}\hspace{-5.5mm}dy\,\sqrt{2\pi R}\,
	m_{\rm BLM}^{\Sigma_{15}}\delta(y-\pi R)\bar\Sigma_{15R}(x,y)\Sigma_{15L}(x)
	+({\rm h.c.})\ ,\\
	\mathcal L_{\rm BLM}^\Delta
 &= \int^{\pi R}_{-\pi R}\hspace{-5.5mm}dy\,\sqrt{2\pi R}\,
	m_{\rm BLM}^\Delta\delta(y)\bar\Delta_L(x,y)\Delta_R(x)
	+({\rm h.c.})\ ,\\
	\mathcal L_{\rm BLM}^\Theta
 &= \int^{\pi R}_{-\pi R}\hspace{-5.5mm}dy\,\sqrt{2\pi R}\,
	m_{\rm BLM}^\Theta\delta(y-\pi R)\bar\Theta_R(x,y)\Theta_L(x)
	+({\rm h.c.})
\intertext{and for three generations $i = 1,2,3$
}
	\mathcal L_{\rm BLM}^Q
 &= \int^{\pi R}_{-\pi R}\hspace{-5.5mm}dy\,
	\sqrt{2\pi R}\,\delta(y)\bar Q_R^i(x)
	\Big\{\eta_{ij}Q^j_{3L}(x,y)+\lambda_{ij}Q^j_L(x,y)\Big\}
	+({\rm h.c.})
\end{align}
\end{subequations}
where
\begin{align}
	Q_L(x,y)
  = \Big[
	\begin{array}{ccc}
	  Q_{6L}^1(x,y) & Q_{6L}^2(x,y) & Q_{15L}(x,y)
	\end{array}
	\Big]^T.
\end{align}
%where 
$Q_R$, $\Sigma_{6,15L}$, $\Delta_R$ and $\Theta_L$
are the brane localized Weyl fermions
of doublet, triplet, quartet and quintet of $SU(2)$ respectively.
The $3\times3$ matrices $\eta_{ij}$, $\lambda_{ij}$ and $m_{\rm BLM}$s are mass parameters.
These brane localized mass terms are introduced at opposite fixed points
such that $Q_R$, $\Delta_R$ ($\Sigma_{6,15L}$, $\Theta_L$) couples
to $Q_{3,6,15L}$, $\Delta_L$ ($\Sigma_{6,15R}$, $\Theta_R$)
localized on the brane at $y=0~(y=\pi R)$.
Let us note that the matrices $\eta_{ij}$, $\lambda_{ij}$ can be non-diagonal,
which are the source of the flavor mixing \cite{2010AKLM, 2011AKLM, BN}.

%%% Flavor mixing %%%%%%%%%%%%%%%%%%%%%%%%%%%%%%%%%%%%%%%%%%%%%%%%%%%%%%%%%%%
\section{Flavor mixing}
In the previous section we worked in the base
where fermion bulk mass terms are written in a diagonal matrix in the generation space.
Then 
%the lagrangian for fermions, which includes 
Yukawa couplings as the gauge interaction of $A_y$ is completely diagonalized
in the generation space.
Thus flavor mixing does not happen in the bulk
and the brane localized mass terms
for the doublets $Q_{3L}$ and $Q_{6L}$\,($Q_{15L}$) is expected to lead to the flavor mixing.
We now 
%confirm the expectation and 
discuss how the flavor mixing is realized in this model.

First, we identify the SM quark doublet
by diagonalizing the relevant brane localized mass term,
%%% \label{diagonal} %%%%%%%%%%%%%%%%%%%%%%
\begin{align}
	\dy\sqrt{2\pi R}\,\delta(y)\bar Q_R(x)
	\big[\begin{array}{ccc}
	 \eta & \lambda
	\end{array}\big]\!\!
	\left[
	\begin{array}{c}
	 Q_{3L}(x,y)\\[3pt]
	 Q_L(x,y)
	\end{array}\right]\!
 &\supset
	\sqrt{2\pi R}\,\bar Q_R(x)
	\big[\begin{array}{ccc}
	 \eta f_L(0) & \lambda f_L(0)
	\end{array}\!\big]\!\!
	\left[
	\begin{array}{c}
	 Q_{3L}(x)\\[3pt]
	 Q_L(x)
	\end{array}\right]\notag\\*
 &= \label{diagonal}
	\sqrt{2\pi R}\, \bar Q_R'(x)
	\big[\begin{array}{ccc}
	 m_{\rm diag} & \bs0_{3\times3}
	\end{array}\!\big]\!\!
	\left[\begin{array}{c}
	 Q_{\H L}(x)\\[1pt]
	 Q_{\SM L}(x)\!\!
	\end{array}\right]
%	\label{masseigenstate}
\end{align}
%%%%%%%%%%%%%%%%%%%%%%%%%%%%%%%%%%%%%%%%%%%
where 
%%% \label{Umatrices} %%%%%%%%%%%%%%%%%%%%%%
\begin{subequations}
\begin{gather}
	\left[\begin{array}{cc}
	 U_1 & U_3\\[3pt]
	 U_2 & U_4
	\end{array}\right]\!\!
	\left[\begin{array}{c}
	 Q_{\H L}(x)\\[3pt]
	 Q_{\SM L}(x)
	\end{array}\right]
  = \left[\begin{array}{c}
	 Q_{3L}(x)\\[3pt]
	 Q_L(x)
	\end{array}\right]~~, \quad
	U^{\bar Q} Q_R(x) 
	%\label{Umatrices} 
  = Q_R'(x)\ ,\\[2pt]
	U^{\bar Q}
	\big[\begin{array}{ccc}
	 \eta f_L(0) & \lambda f_L(0)
	\end{array}\!\big]\!\!
	\left[
	\begin{array}{cc}
	 U_1 & U_3\\[3pt]
	 U_2 & U_4
	\end{array}
	\right] 
  = \big[\begin{array}{ccc}
	 m_{\rm diag} & \bs0_{3\times3}
	\end{array}\!\big]\ .
\end{gather}
\end{subequations}
%%%%%%%%%%%%%%%%%%%%%%%%%%%%%%%%%%%%%%%%%%%
In eq.\,\eqref{diagonal},
$\eta f_L(0)$ is an abbreviation of a $3\times3$ matrix
whose $(i, j)$ element is given by $\eta_{ij}f_L^j(0)$, for instance.
$U_3$, $U_4$ are $3\times3$ matrices satisfying the unitarity condition 
\begin{equation}
\label{unitary_cond}
U_3^\dag U_3+U_4^\dag U_4={\bf 1}_{3\times3}, 
\end{equation}
which indicates how the quark doublets of the SM are contained
in each of $ Q_{3L}(x)$ and $Q_{6,15L}(x)$
and compose a $6\times6$ unitary matrix together with $U_1$, $U_2$,
which diagonalizes the brane localized mass matrix.
The eigenstate $Q_\H$ becomes massive and decouples from the low energy processes,
while $Q_\SM$ remains massless at this stage
and is identified with the SM quark doublet.
%$U_3$ and $U_4$ satisfy the following unitarity condition:
%%% \label{unitary_cond} %%%%%%%%%%%%%%%%%%%%%%
%\begin{equation}
%\label{unitary_cond} 
%  U_3^\dag U_3+U_4^\dag U_4={\bf 1}_{3\times3}\ .
%\end{equation} 
%%%%%%%%%%%%%%%%%%%%%%%%%%%%%%%%%%%%%%%%%%%
After this identification of the SM doublet,
Yukawa couplings are read off from the higher dimensional gauge interaction of $A_y$,
whose zero mode is the Higgs field $H(x)$:
\begin{align}
	-\frac{g_4}2\!
	\left\{
	\left\langle H^{\dagger}\right\rangle\bar d_R^i(x)
	I_{RL}^{i(00)}U_3^{ij}Q_{\SM L}^j(x)
	+\left\langle H^t\right\rangle i\sigma^2
	 \bar u_R^i(x)\big(WI_{RL}^{(00)}\big)^iU_4^{ij}Q_{\SM L}^j(x)
	\right\}
	+{\rm h.c.}
\end{align}
where $g_4\equiv \frac g{\sqrt{2\pi R}}$
and the matrix $W$ indicates the factor $\sqrt n$ in \eqref{rankfactor}:
\begin{align}
	W \equiv {\rm diag}\Big(\sqrt2\,, \sqrt2\,, 2\Big)
\end{align}
and $I_{RL}^{(00)}$ is an overlap integral of mode functions of fermions 
with matrix elements $\big(I_{RL}^{(00)}\big)_{ij} = \delta_{ij} I_{RL}^{i(00)}$:
\begin{align}
	I_{RL}^{i(00)}
 = \int^{\pi R}_{-\pi R}\hspace{-5.5mm}dy\,f_L^if_R^i
 = \left\{
	\begin{array}{cl}
	 \dis\frac{\pi RM^i}{\sinh(\pi RM^i)} &\quad \big(i = 1,2\big) \\[12pt]
	 1 &\quad \big(i = 3\big)
	\end{array}
	\right. 
\end{align}
which behaves as $2\pi RM^i\e^{-\pi RM^i}$ for $\pi RM^i \gg 1$,
thus realizing the hierarchical small quark masses without fine tuning of $M^i$.
We thus know that the matrices of Yukawa coupling
$\frac{g_4}2Y_u$ and $\frac{g_4}2Y_d$ are given as
%%% \label{Yukawa coupling} %%%%%%%%%%%%%%%%%%%%%%
\begin{align} 
	\label{Yukawa coupling} 
	\frac{g_4}2Y_u = \frac{g_4}2WI_{RL}^{(00)} U_4\ , \qquad
	\frac{g_4}2Y_d = \frac{g_4}2I_{RL}^{(00)} U_3.
\end{align}
%where 
%the matrix $I_{RL}^{(00)}$ has elements
%$\big(I_{RL}^{(00)}\big)_{ij} = \delta_{ij} I_{RL}^{i(00)}$.
These matrices are diagonalized by bi-unitary transformations as in the SM
and Cabibbo-Kobayashi-Maskawa (CKM) matrix is defined in a usual way \cite{CKMmatrix}.
%%% \label{cond_U3,U4} %%%%%%%%%%%%%%%%%%%%%%
\begin{align}
	\label{cond_U3,U4}
	\left\{
	\begin{aligned}
	  \hat Y_d &= {\rm diag}(\hat m_d,\cdots) = V_{dR}^\dag Y_d V_{dL}\\
	  \hat Y_u &= {\rm diag}(\hat m_u,\cdots) = V_{uR}^\dag WY_u V_{uL}
	\end{aligned}
	\right.\quad , \qquad
	  V_{\rm CKM}\equiv V_{dL}^\dag V_{uL}
\end{align}
where all the quark masses are normalized
by the $W$-boson mass as $\hat m_f =\frac{m_f}{M_W}$.
A remarkable point is that the Yukawa couplings $\frac{g_4}2Y_u$ and $\frac{g_4}2Y_d$ are
related through the unitarity condition eq.\,\eqref{unitary_cond},
on the contrary those are completely independent in the SM.

For an illustrative purpose to confirm the mechanism of flavor mixing,
we will see how the realistic quark masses and mixing are reproduced.
Here we leave aside CP violation since the issue discussed in this paper is independent of it 
and assume that $U_3$, $U_4$ are real. 
Let us notice that $3\times3$ matrices $U_{3,4}$ can be parametrized
because of \eqref{unitary_cond} without loss of generality as
\begin{align} 
%%% \label{parametrization} %%%%%%%%%%%%%%%%%%%%%%
	\label{parametrization}
	U_4
  = R_u\!
	\left[
	\begin{array}{ccc}
	 a_1 & 0 & 0\\[3pt]
	 0 & a_2 & 0\\[3pt]
	 0 & 0 & a_3
	\end{array}
	\right]\quad, \qquad 
	U_3
  = R_d\!
	\left[
	\begin{array}{ccc}
	 \sqrt{1-a_1^2} & 0 & 0\\[2pt]
	 0 & \sqrt{1-a_2^2} & 0\\[2pt]
	 0 & 0 & \sqrt{1-a_3^2}
	\end{array}
	\right]
\end{align}
where $R_u$ and $R_d$ are arbitrary $3\times3$ rotation matrices parametrized as
\begin{subequations}
\begin{align}
	R_u %\label{3gparameterizationRu}
 &= \left[\begin{array}{ccc}
	  1	& 0	& 0\\[3pt]
	  0	& \cos\theta'_2 & \sin\theta'_2\\[3pt]
	  0	& -\!\sin\theta'_2 & \cos\theta'_2
	\end{array}\right]\!\!\!
	\left[\begin{array}{ccc}
	  \cos\theta'_3 & 0 & \sin\theta'_3\\[3pt]
	  0 & 1 & 0\\[3pt]
	  -\!\sin\theta'_3 & 0 & \cos\theta'_3
	\end{array}\right]\!\!\!
	\left[\begin{array}{ccc}
	  \cos\theta'_1 & -\!\sin\theta'_1 & 0\\[3pt]
	  \sin\theta'_1 & \cos\theta'_1 & 0\\[3pt]
	  0 & 0 & 1
	\end{array}\right]\ ,\\
	R_d
 &= \left[\begin{array}{ccc}
	  1	& 0	& 0\\[3pt]
	  0	& \cos\theta_2 & \sin\theta_2\\[3pt]
	  0	& -\!\sin\theta_2 & \cos\theta_2
	\end{array}\right]\!\!\!
	\left[\begin{array}{ccc}
	  \cos\theta_3 & 0 & \sin\theta_3\\[3pt]
	  0 & 1 & 0\\[3pt]
	  -\!\sin\theta_3 & 0 & \cos\theta_3
	\end{array}\right]\!\!\!
	\left[\begin{array}{ccc}
	  \cos\theta_1 & -\!\sin\theta_1 & 0\\[3pt]
	  \sin\theta_1 & \cos\theta_1 & 0\\[3pt]
	  0 & 0 & 1
	\end{array}\right]\ .
\end{align}
\end{subequations}
Actually the most general forms of $U_3$ and $U_4$ have a common orthogonal matrix
multiplied from the right,
which can be however eliminated
by suitable unitary transformations among the members of $Q_{\SM L}(x)$.

Now physical observables
$\hat m_u$, $\hat m_c$, $\hat m_t$, $\hat m_d$, $\hat m_s$, $\hat m_b$
and the angles of the CKM matrix are expressed 
in terms of $a_i$, $b_i\big(\!\!\equiv I_{RL}^{i(00)}\big)$
and 6 rotation angles in $R_u$ and $R_d$.
Note that our theory has 2 free parameters which cannot be determined by the observables 
since 9 physical observables are written in terms of 11 parameters.

As we have discussed in the previous paper \cite{2011AKLM},
if the large mixings between the 1-3 and 2-3 generations are introduced then the top quark mass decreases from 160 GeV $\sim 2M_W$.
Thus, we expect that the mixing angles between the third generation and the first two generations are considered to be small to keep $m_t \sim 2M_W$. 
Also the relation between the masses of top and bottom quark $m_t^2+(2m_b)^2=(2M_W)^2$ for the $M_3=0$ holds and we must choose $a_3 \sim 1$. 
%The mixing angles between the third generation and the first two generations are considered to be small 
%because the mass differences among them are large. 
%in order to keep the top mass $m_t \sim 2M_W$.
It implies that the rotation angles
$\theta_2'$, $\theta_3'$, $\theta_2$, $\theta_3$
and parameter $\sqrt{1-a_3^2}$ should be small,
and also other 6 parameters $a_1$, $a_2$, $b_1$, $b_2$ and $\theta_1'$, $\theta_1$
should take values close to those of 2 generation model\cite{2011AKLM}.

Actually, for the case of $R_u = \bs1_{3\times3}$ where the up type quark mixings vanish,
this case gives almost the most stringent lower bound from $K^0$\,--\,$\bar K^0$ mixing 
for example,
these parameters are numerically found as
\begin{alignat}{5}
	a_1^2 \approx 0.1023 &\qquad&
	b_1^2 \approx 4.355\times10^{-9} &\qquad&&
   \sin\theta_1 \approx -2.587\times10^{-2}\notag\\
	a_2^2 \approx 0.9887 &\quad , \qquad&
	b_2^2 \approx 1.302\times10^{-4} &\quad , \qquad&&
   \sin\theta_2 \approx 2.224\times10^{-2}~~.\\
	a_3^2 \approx 0.9966 &\qquad&
	&\qquad&&
   \sin\theta_3 \approx 2.112\times10^{-4}\notag
\end{alignat}
Also, for the another case of $R_d = \bs1_{3\times3}$ where the down type quark mixings vanish,
these parameters are numerically found as
\begin{alignat}{5}
	a_1^2 \approx 0.0650 &\qquad&
	b_1^2 \approx 3.973\times10^{-9} &\qquad&&
   \sin\theta_1' \approx 0.6704\notag\\
	a_2^2 \approx 0.9931 &\quad , \qquad&
	b_2^2 \approx 2.235\times10^{-4} &\quad , \qquad&&
   \sin\theta_2' \approx -3.936\times10^{-2}~~.\\
	a_3^2 \approx 0.9966 &\qquad&
	&\qquad&&
   \sin\theta_3' \approx 1.773\times10^{-2}\notag
\end{alignat}
These two results show that the mixing angles $\theta_2,\theta_3,\theta_2',\theta_3'\ll 1$ and it is completely consistent with the above argument.

%%% B^0-\bar B^0 mixing %%%%%%%%%%%%%%%%%%%%%%%%%%%%%%%%%%%%%%%%%%%%%%%%%%%%%%%%% 
\section[$B^0$\,--\,$\bar B^0$ mixing]{$\bs{B^0}$\,--\,$\bs{\bar B^0}$ mixing}
In this section,
we apply the results of the previous section to representative FCNC processes,
$B_d^0$\,--\,$\bar B_d^0$ mixing and $B_s^0$\,--\,$\bar B_s^0$ mixing
responsible for the mass difference of two neutral $B$ mesons.%
\footnote{%
For the studies of $B^0$\,--\,$\bar B^0$ mixing in other new physics models,
see for instance \cite{2011HKMRS}
}

We focus on the FCNC processes of zero-mode down-type quarks
due to gauge boson exchange at the tree level.
First let us consider the processes with the exchange of zero mode gauge bosons.
If such type of diagrams exist with a sizable magnitude,
it will easily spoil the viability of the model.

Concerning the $Z$-boson exchange,
it is in principle possible to occur the tree-level FCNC.
Since the mode function of the zero-mode gauge boson is $y$-\hspace{0pt}independent,
the overlap integral of mode functions is 
%universal, i.e. 
generation independent. 
%just as the kinetic term of fermions are.
Thus the gauge coupling of zero mode gauge boson depends on only the relevant quantum numbers
such as the third component of weak isospin $I_3$.
Therefore the condition proposed by Glashow-Weinberg \cite{GW}
to guarantee natural flavor conservation for the theories of 4D space-time is relevant.

Although there are right-handed down-type quarks belonging to different representations
in our model, for example, the $SU(2)$ singlet $d_R$ in $\psi(\bs3)$
and one of components of the triplet $\Sigma_R$
in $\psi(\bar{\bs6})$ or $\psi(\overline{\bf15})$,
these are known to have the same quantum number $I_3 = 0$,
and thus the Glashow-Weinberg condition is satisfied in this sector \cite{2010AKLM}.
However, the quintet $\Theta_R$ in $\psi(\overline{\bf15})$ also contains %$d_R$
the right-handed down-type quark,
and this has the different quantum number $I_3 = 1$
from that of $d_R^i$ belonging to $\psi(\bs3)$. 
%while they have the same electric charge and chirality.

%Even more unfortunately,
What is worse, 
the quartet $\Delta_L$ in $\psi(\overline{\bf15})$ contains left-handed down-type quark 
%and this has 
with the different quantum number $I_3 = \frac12$ from that of $d_L^i$
belonging to the doublet $Q_L$
in $\psi(\bs3)$, $\psi(\bar{\bs6})$ or $\psi(\overline{\bf15})$ 
%which have 
with the quantum number $I_3 = -\frac12$.
Thus, the condition of Glashow-Weinberg is not satisfied
in the down-type quark sector and FCNC process
due to the exchange of the zero mode $Z$-boson arises at the tree level.%
\footnote{%
The FCNC due to the exchanges of zero-mode photon and gluon trivially vanish 
because the fermions of our interest have the same electric charge and color.
}
However, the quintet $\Theta_R$ (quartet $\Delta_L$) is an exotic fermion
and acquires large $SU(2)$ invariant brane mass.
Thus the mixing between $d_R^i$ ($d_L^i$) and $\Theta_R$ ($\Delta_L$) is inversely suppressed
by the power of $m_{\rm BLM}$ and the FCNC vertex of $Z$-boson can be safely neglected.
We may say that the condition of Glashow-Weinberg is satisfied in a good approximation
in the processes via the zero mode gauge boson exchange.
%Furthermore, the contribution by the weak gauge boson exchange is expected
%to be small compared with that by the gluon exchange.

One may worry that $Z'$ gauge boson exchange give rise to FCNC processes 
since an extra $U(1)$ gauge symmetry is indispensable for getting a realistic Weinberg angle. 
Note that the extra $U(1)$ gauge symmetry is explicitly broken by an anomaly and 
the gauge boson of the extra $U(1)$ gauge symmetry acquire a mass of the cutoff scale order. 
In our model, the cutoff scale is a 5D Planck scale which is larger than the intermediate scale $10^{13}$ GeV. 
Therefore, the FCNC effects by $Z'$ gauge boson exchange can be safely neglected 
comparing to the process by non-zero KK gluon exchanges considered later. 

Hence, the remaining possibility is the process via the exchange of non-zero KK gauge bosons.
In this case, the mode functions of KK gauge bosons are $y$-\hspace{0pt}dependent
and their couplings to fermions are no longer universal  
because of non-degenerate bulk masses, even if the condition of Glashow-Weinberg is met.

Therefore, such progresses lead to FCNC at the tree level.
In our previous papers \cite{2010AKLM, 2011AKLM},
we have calculated $K^0$\,--\,$\bar K^0$ mixing and $D^0$\,--\,$\bar D^0$ mixing
via the non-zero KK gluon exchange at the tree level
and obtained a lower bound of the compactification scale as the prediction of our model.
Along the same line of the argument as in our previous papers,
we here study $B_d^0$\,--\,$\bar B_d^0$ and $B_s^0$\,--\,$\bar B_s^0$ mixings 
in the down-type quark sector by the non-zero KK gluon exchange at the tree level
as the dominant contribution to these FCNC processes.

For such purpose, let us derive the four dimensional effective QCD interaction vertices 
for the zero modes of down-type quarks relevant for our calculation:
%%% \label{strongc} %%%%%%%%%%%%%%%%%%%%%%
\begin{align}
	\mathcal L_\s
 &\supset
 	\frac{g_\s}{2\sqrt{2\pi R}}G_\mu^a\!
	\left(
	\bar{d}_R^i\lambda^a\gamma^\mu d_R^i
	+\bar{d}_L^i\lambda^a\gamma^\mu d_L^i
	\right)
	+\frac{g_\s}2G_\mu^{a(n)}
	\bar{d}_R^i\lambda^a\gamma^\mu d_R^j\!
	\left(V_{dR}^\dagger I_{RR}^{(0n0)}V_{dR}\right)_{ij}
	\notag\\*
 &\hspace{4mm} \label{strongc}
	+\!\frac{g_\s}2G_\mu^{a(n)}
	\bar{d}_L^i\lambda^a\gamma^\mu d_L^j(-1)^n\!
	\left\{
	V_{dL}^\dagger\!
	\left(
	U_3^\dagger I_{RR}^{(0n0)}U_3+U_4^\dagger I_{RR}^{(0n0)}U_4
	\right)\!
	V_{dL}
	\right\}_{ij}\ .
\end{align}
%%%%%%%%%%%%%%%%%%%%%%%%%%%%%%%%%%%%%%%%%%%%%%%%%%%%
where $I_{RR}^{i(0n0)}$ and $I_{LL}^{i(0n0)}$ are overlap integrals relevant for gauge interaction,
%%% \label{vertexfunction} %%%%%%%%%%%%%%%%%%%%%%
%\begin{equation} 
\begin{eqnarray}
%   \label{vertexfunction} 
   &&I_{RR}^{i(0n0)}
 = \frac1{\sqrt{\pi R}}\dy\big(f^i_R\big)^2\cos\frac nRy
 = \frac1{\sqrt{\pi R}}
   \frac{(2RM^i)^2}{(2RM^i)^2+n^2}
   \frac{(-1)^n\e^{2\pi RM^i}-1}{\e^{2\pi RM^i}-1}, \\   
%\end{equation}
%%%%%%%%%%%%%%%%%%%%%%%%%%%%%%%%%%%%%%%%%%%%%%%%%%%%
%Let us note that
%the overlap integrals for left-handed fermion $I_{LL}^{i(0n0)}$ are related
%to those for the right-handed ones $I_{RR}^{i(0n0)}$ as
%\begin{align}
	&&I_{LL}^{i(0n0)}
 = I_{RR}^{i(0n0)}\Big|_{M^i\,\to\, -M^i}
  = (-1)^nI_{RR}^{i(0n0)}
\end{eqnarray}
%\end{align}
since the chirality exchange corresponds to the exchange of two fixed points.
%In eq.\,\eqref{strongc},
%$\tilde d$ denotes mass eigenstates, $\big(\tilde d^1, \tilde d^2\big) = (d, s)$.
We can see from \eqref{strongc} that the FCNC appears in the couplings of non-zero KK gluons
due to the fact that $I_{RR}^{(0n0)}$ is not proportional
to the unit matrix in the generation space,
while the coupling of the zero mode gluon is flavor conserving, as we expected.

\begin{figure}[t]
\[
\begin{array}{ccc}
\includegraphics[bb= 0 0 106 66]{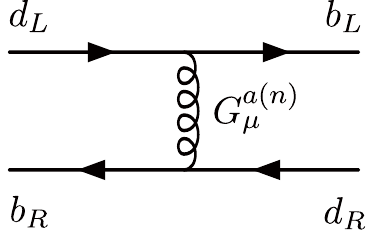}
	\qquad & \qquad
\includegraphics[bb= 0 0 106 66]{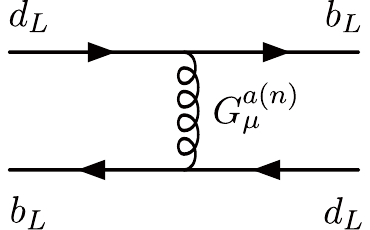}
	\qquad & \qquad
\includegraphics[bb= 0 0 106 66]{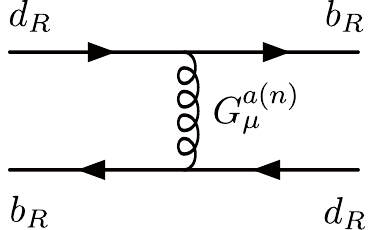}\\[8pt]
\text{(i) LR type} \qquad&\qquad
\text{(ii) LL type} \qquad&\qquad
\text{(iii) RR type}%\\[25pt]
\end{array}
\]
\caption{\small%
The diagrams of $B_d^0$\,--\,$\bar B_d^0$ %(upper line) and 
mixing via KK gluon exchange. 
Those of $B_s^0$\,--\,$\bar B_s^0$ %(bottom line) 
mixing via KK gluon exchange are obtained by the replacements $d \leftrightarrow s$.}
\label{fig1}
\end{figure}

The Feynman rules necessary for the calculation of $B_d^0$\,--\,$\bar B_d^0$ mixing %and $B_s^0$\,--\,$\bar B_s^0$ mixings 
can be read off from \eqref{strongc}.
\begin{subequations}
\begin{align}
	\begin{array}{c}
	\includegraphics[bb= 0 0 169 59, scale=0.65]{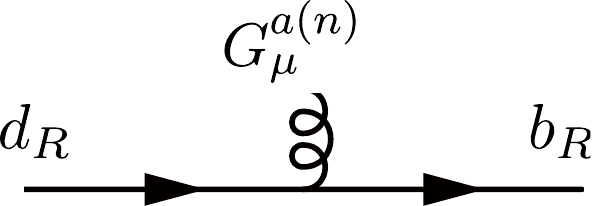}~
	\end{array}
 &= \frac{g_\s}2\!
	\left(V_{dR}^\dagger I_{RR}^{(0n0)}V_{dR}\right)_{31}\!\!
	\lambda^a\gamma^\mu R\ ,\\[5pt]
%%% \label{BBvertexLL} %%%%%%%%%%%%%%%%%%%%%%
	\label{BBvertexLL}
	\begin{array}{c}
	\includegraphics[bb= 0 0 169 59, scale=0.65]{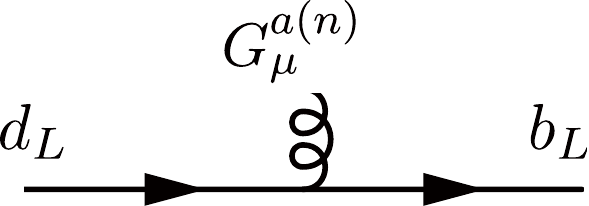}~
	\end{array}
 &= \frac{g_\s}2(-1)^n\!
	\left\{
	V_{dL}^\dagger\!
	\left(
	U_3^\dagger I_{RR}^{(0n0)}U_3+U_4^\dagger I_{RR}^{(0n0)}U_4
	\right)\!
	V_{dL}
	\right\}_{31}\!\!
	\lambda^a\gamma^\mu L\ ,\\[5pt]
	\begin{array}{c}
	\includegraphics[bb= 0 0 119 31]{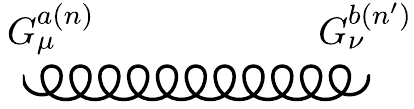}
	\end{array}
 &= \delta_{nn'}\delta_{ab}\frac{\eta^{\mu\nu}}{k^2-M_n^2}
	\qquad\Big(\,\text{'t Hooft-Feynman gauge}\,\Big)\ .
%%%%%%%%%%%%%%%%%%%%%%%%%%%%%%%%%%%%%%%%%%%%%%%%%%%%
\end{align}
\end{subequations}
Those for $B_s^0$\,--\,$\bar B_s^0$ mixing are easily obtained by the replacements $d \leftrightarrow s$ 
and $31 \leftrightarrow 32$ in the matrix element of the vertices. 
The non-zero KK gluon exchange diagrams providing the dominant contribution
to the process of $B_d^0$\,--\,$\bar B_d^0$ and $B_s^0$\,--\,$\bar B_s^0$ mixing 
are depicted in Fig.\,\ref{fig1}.
%%% figure 1 %%%%%%%%%%%%%%%%%%%%%%
%%% figure 1 %%%%%%%%%%%%%%%%%%%%%%

By noting the fact $k^2\ll (\frac nR)^2$ for $n \neq 0$
being the mass of $n$-th KK gluon and $k^\mu$ being internal momentum,
the contributions from each type diagram 
of the $B_d^0$\,--\,$\bar B_d^0$ mixing in Fig.\,\ref{fig1} are written
in the form of effective four-Fermi lagrangian obtained by use of Feynman rules listed above,
\begin{subequations}
%%% \label{BBbarLR} %%%%%%%%%%%%%%%%%%%%%%
	\label{BBbar}
%%% \label{BBbarLR} %%%%%%%%%%%%%%%%%%%%%%
%%% \label{BBbarLL} %%%%%%%%%%%%%%%%%%%%%%
%%% \label{BBbarRR} %%%%%%%%%%%%%%%%%%%%%%
\begin{align}
	\begin{array}{c}
	\includegraphics[bb= 0 0 106 66]{./BBfigure/Bd-BdLR1.pdf}
	\end{array}
 \sim &
	-\sum^\infty_{n=1}\frac{g_\s^2}4\frac1{M_c^2}\!
	\left(\bar b_L\lambda^a\gamma^\mu d_L\right)\!
	\left(\bar b_R\lambda^a\gamma_\mu d_R\right) \times \notag\\[-10pt]
  &\,%\label{BBbarLR} 
	\frac{(-1)^n}{n^2}\!
	\left\{
	V_{dL}^\dagger\!
	\left(
	U_3^\dagger I_{RR}^{(0n0)}U_3+U_4^\dagger I_{RR}^{(0n0)}U_4
	\right)\!
	V_{dL}
	\right\}_{31}\!\!
	\left(V_{dR}^\dagger I_{RR}^{(0n0)}V_{dR}\right)_{31}, \\
	\begin{array}{c}
	\includegraphics[bb= 0 0 106 66]{./BBfigure/Bd-BdLL2.pdf}
	\end{array}
 \sim &
	-\sum_{n=1}^\infty
	\frac{g_\s^2}4\frac1{M_c^2}\!
	\left(\bar b_L\lambda^a\gamma^\mu d_L\right)\!
	\left(\bar b_L\lambda^a\gamma_\mu d_L\right) \times \notag\\*[-10pt]
 &\,%\label{BBbarLL} 
	\frac1{n^2}\!
	\left\{
	V_{dL}^\dagger\!
	\left(
	U_3^\dagger I_{RR}^{(0n0)}U_3+U_4^\dagger I_{RR}^{(0n0)}U_4
	\right)\!
	V_{dL}
	\right\}_{31}^2, \\[0.5\bls]
	\begin{array}{c}
	\includegraphics[bb= 0 0 106 66]{./BBfigure/Bd-BdRR3.pdf}
	\end{array}
 \sim & %\label{BBbarRR}
	-\sum_{n=1}^\infty
	\frac{g_\s^2}4\frac1{M_c^2} \frac{1}{n^2} \!
	\left(\bar b_R\lambda^a\gamma^\mu d_R\right)\!
	\left(\bar b_R\lambda^a\gamma_\mu d_R\right)\!\!
	\left(V_{dR}^\dagger I_{RR}^{(0n0)}V_{dR}\right)_{31}^2. 
\end{align} 
%%%%%%%%%%%%%%%%%%%%%%%%%%%%%%%%%%%%%%%%%%%%%%%%%%%%
\end{subequations}
Similarly, those for $B_s^0$\,--\,$\bar B_s^0$ mixing are obtained 
by the replacements $d \leftrightarrow s$ and $31 \leftrightarrow 32$ in the matrix element of the vertices. 
The sum over the integer $n$ is convergent and the coefficients of the effective lagrangian
\eqref{BBbar} for the $B_d^0$\,--\,$\bar B_d^0$ mixing
and \eqref{BBbar} after the replacements of $d \leftrightarrow s$ and  $31 \leftrightarrow 32$ 
for the $B_s^0$\,--\,$\bar B_s^0$ mixing
are suppressed by the compactification scale as $1/M_{\rm c}^2=R^2$. %where $M_{\rm c} = R^{-1}$.
%We can verify, as we expect, that the coefficient vanishes
%in the limit of universal bulk masses
%$M^1 = M^2 = M^3$ by use of the unitarity condition \eqref{unitary_cond},
%since $I_{RR}^{(0n0)}$ is proportional to the unit matrix in this limit;
%\begin{align}
%	V_{uL}^\dag\!
%	\left(
%	U_3^\dag I_{RR}^{(0m0)}U_3
%	+U_4^\dag I_{RR}^{(0m0)}U_4
%	\right)\! V_{uL}
% &~\xrightarrow{M^1 = M^2 = M^3\,}~
%	V_{uL}^\dag\!
%	\left(U_3^\dag U_3+U_4^\dag U_4\right)\!
%	V_{uL}I_{RR}^{(0m0)}
% \propto {\mathbf 1}_{3\times3}\ ,\notag\\
%
%	V_{uR}^\dag I_{RR}^{(0m0)}V_{uR}
% &~\xrightarrow{M^1 = M^2 = M^3\,}~
%	V_{uR}V_{uR}^\dag I_{RR}^{(0m0)}
%  \propto {\mathbf 1}_{3\times3}\ .
%\end{align}

Comparing the results %of \eqref{BBbar} and \eqref{BsBsbar} 
with the experimental data,
we can estimate a lower bound on the compactification scale.
The most general effective Hamiltonian for $\Delta B=2$ processes
due to some \lq\lq new physics" at a high scale $\Lambda_{\rm NP} \gg M_W$
can be written as follows;
%%% \label{effectiveH} %%%%%%%%%%%%%%%%%%%%%%
\begin{align} 
%	\label{effectiveH}
	\mathcal H_{\rm eff}^{\Delta B=2}
  = \frac1{\Lambda_{\rm NP}^2}\!
	\left(
	\sum_{i=1}^5z^q_iQ^q_i+\sum_{i=1}^3 \tilde z^q_i\tilde Q^q_i
	\right)
\end{align}
%%%%%%%%%%%%%%%%%%%%%%%%%%%%%%%%%%%%%%%%%%%%%%%%%%%%
where the 4-Fermi operators relevant for $B_d^0$\,--\,$\bar B_d^0$ mixing are given as,
\begin{subequations}
%%% \label{Qi} %%%%%%%%%%%%%%%%%%%%%%
	\label{Qi}
%%% \label{QiBd} %%%%%%%%%%%%%%%%%%%%%%
\begin{gather}
	Q_1^d
  = \bar d_L^\alpha \gamma_\mu b_L^\alpha\bar d_L^\beta \gamma^\mu b^\beta_L\ ,\qquad
	Q_2^d
  = \bar d_R^\alpha b_L^\alpha\bar d_R^\beta b^\beta_L\ ,\qquad
	Q_3^d
  = \bar d_R^\alpha b_L^\beta\bar d_R^\beta b^\alpha_L\ ,\notag\\*
%	\label{QiBd}
	Q_4^d
  = \bar d_R^\alpha b_L^\alpha\bar d_L^\beta b^\beta_R\ ,\qquad
	Q_5^d
  = \bar d_R^\alpha b_L^\beta\bar d_L^\beta b^\alpha_R
\end{gather}
%%%%%%%%%%%%%%%%%%%%%%%%%%%%%%%%%%%%%%%%%%%%%%%%%%%%
and for $B_s^0$\,--\,$\bar B_s^0$ mixing,
%%% \label{QiBd} %%%%%%%%%%%%%%%%%%%%%%
\begin{gather}
	Q_1^s
  = \bar s_L^\alpha \gamma_\mu b_L^\alpha\bar s_L^\beta \gamma^\mu b^\beta_L\ ,\qquad
	Q_2^s
  = \bar s_R^\alpha b_L^\alpha\bar s_R^\beta b^\beta_L\ ,\qquad
	Q_3^s
  = \bar s_R^\alpha b_L^\beta\bar s_R^\beta b^\alpha_L\ ,\notag\\*
%	\label{QiBd}
	Q_4^s
  = \bar s_R^\alpha b_L^\alpha\bar s_L^\beta b^\beta_R\ ,\qquad
	Q_5^s
  = \bar s_R^\alpha b_L^\beta\bar s_L^\beta b^\alpha_R. 
\end{gather}
\end{subequations}
%%%%%%%%%%%%%%%%%%%%%%%%%%%%%%%%%%%%%%%%%%%%%%%%%%%%
%%%%%%%%%%%%%%%%%%%%%%%%%%%%%%%%%%%%%%%%%%%%%%%%%%%%
Indices $\alpha$, $\beta$ stand for the color degrees of freedom.
The operators $\tilde Q_{1,2,3}$ are obtained from the $Q_{1,2,3}$
by the chirality exchange $L \leftrightarrow R$.
Since the SM contribution is poorly known,
we can get the constraint on the new physics directly from the experimental data
assuming that there is no accidental cancellation
between the contributions of the SM and new physics.
If we assume one of these possible operators gives dominant contribution to the mixing,
each coefficient is independently constrained as follows,
with the same constraints for $\tilde z^q_i$ as those for $z^q_i$ $(i = 1,2,3)$ \cite{UTfit};
%%% \label{BBconstraints} %%%%%%%%%%%%%%%%%%%%%%
\begin{align}
\label{BBconstraints}
\begin{array}{cccccc}
   |z^d_1|\dis \leq 2.3\times10^{-5}\left(\frac{\Lambda_{\rm NP}}{\rm 1TeV}\right)^2
 & \qquad\qquad\qquad
 & |z^s_1|\dis \leq 1.1\times10^{-3}\left(\frac{\Lambda_{\rm NP}}{\rm 1TeV}\right)^2\\[9pt]
   |z^d_2|\dis \leq 7.2\times10^{-7}\left(\frac{\Lambda_{\rm NP}}{\rm 1TeV}\right)^2 &
 & |z^s_2|\dis \leq 5.6\times10^{-5}\left(\frac{\Lambda_{\rm NP}}{\rm 1TeV}\right)^2\\[9pt]
   |z^d_3|\dis \leq 2.8\times10^{-6}\left(\frac{\Lambda_{\rm NP}}{\rm 1TeV}\right)^2 &,
 & |z^s_3|\dis \leq 2.1\times10^{-4}\left(\frac{\Lambda_{\rm NP}}{\rm 1TeV}\right)^2\\[9pt]
   |z^d_4|\dis \leq 2.1\times10^{-7}\left(\frac{\Lambda_{\rm NP}}{\rm 1TeV}\right)^2 &
 & |z^s_4|\dis \leq 1.6\times10^{-5}\left(\frac{\Lambda_{\rm NP}}{\rm 1TeV}\right)^2\\[9pt]
   |z^d_5|\dis \leq 6.0\times10^{-7}\left(\frac{\Lambda_{\rm NP}}{\rm 1TeV}\right)^2 &
 & |z^s_5|\dis \leq 4.5\times10^{-5}\left(\frac{\Lambda_{\rm NP}}{\rm 1TeV}\right)^2
\end{array}
\end{align}
%%%%%%%%%%%%%%%%%%%%%%%%%%%%%%%%%%%%%%%%%%%%%%%%%%%%
where the new physics scale $\Lambda_{\rm NP}$ is regarded as the compactification scale in our case.
All we have to do is to represent \eqref{BBbar} and its replacements $d \leftrightarrow s$ and $31 \leftrightarrow 32$ of \eqref{BBbar}
by use of \eqref{Qi} and to utilize these constraints \eqref{BBconstraints}.

We can rewrite the each type effective lagrangian
for $B_d^0$\,--\,$\bar B_d^0$ mixing \eqref{BBbar}
in terms of effective Hamiltonian by using the Fierz transformation 
and the completeness condition for Gell-Mann matrices;
%%% \label{BBbar2} %%%%%%%%%%%%%%%%%%%%%%
\begin{align} \label{BBbar2}
	\mathcal H^{\Delta B=2}_{{\rm eff}, LL}
 &= \frac{z^d_1Q^d_1}{R^{-2}}\quad , \quad
	\mathcal H^{\Delta B=2}_{{\rm eff}, RR}
  = \frac{\tilde z^d_1\tilde Q^d_1}{R^{-2}}\quad , \quad
	\mathcal H^{\Delta B=2}_{{\rm eff}, LR}
  = \frac{z^d_4Q^d_4+z^d_5Q^d_5}{R^{-2}}\ 
\end{align}
%%%%%%%%%%%%%%%%%%%%%%%%%%%%%%%%%%%%%%%%%%%%%%%%%%%%
where
%%% \label{BBbar2z} %%%%%%%%%%%%%%%%%%%%%%
\begin{subequations} \label{BBbar2z}
\begin{align}
	z_1^d
 &= \frac{8\pi\alpha_\s}3
	\pi R\sum^\infty_{n=1}\frac1{n^2}\!
	\left\{
	V_{dL}^\dagger\!
	\left(
	U_3^\dagger I_{RR}^{(0n0)}U_3+U_4^\dagger I_{RR}^{(0n0)}U_4
	\right)\!
	V_{dL}
	\right\}_{31}^2\ ,\\
	\tilde z_1^d
 &= \frac{8\pi\alpha_\s}3
	\pi R\sum^\infty_{n=1}\frac1{n^2}\!
	\left(V_{dR}^\dagger I_{RR}^{(0n0)}V_{dR}\right)_{31}^2\ ,\\
%%% \label{BBbar2zLR4} %%%%%%%%%%%%%%%%%%%%%%
	\label{BBbar2zLR4}
	z_4^d
 &= -8\pi\alpha_\s
	\pi R\sum^\infty_{n=1}
	\frac{(-1)^n}{n^2}\!
	\left\{
	V_{dL}^\dagger\!
	\left(
	U_3^\dagger I_{RR}^{(0n0)}U_3+U_4^\dagger I_{RR}^{(0n0)}U_4
	\right)\!
	V_{dL}
	\right\}_{31}\!\!
	\left(V_{dR}^\dagger I_{RR}^{(0n0)}V_{dR}\right)_{31}\!,\\
%%% \label{BBbar2zLR5} %%%%%%%%%%%%%%%%%%%%%%
	\label{BBbar2zLR5}
	z_5^d
 &= \frac83\pi\alpha_\s
	\pi R\sum^\infty_{n=1}
	\frac{(-1)^n}{n^2}\!
	\left\{
	V_{dL}^\dagger\!
	\left(
	U_3^\dagger I_{RR}^{(0n0)}U_3+U_4^\dagger I_{RR}^{(0n0)}U_4
	\right)\!
	V_{dL}
	\right\}_{31}\!\!
	\left(V_{dR}^\dagger I_{RR}^{(0n0)}V_{dR}\right)_{31}\!.
\end{align}
\end{subequations}
%%%%%%%%%%%%%%%%%%%%%%%%%%%%%%%%%%%%%%%%%%%%%%%%%%%%
The four-dimensional $\alpha_\s$ is defined by
\begin{equation}
	\alpha_\s
  = \frac{\big(g_\s^{4D}\big)^2}{4\pi}
  = \frac1{2\pi R}\frac{g_\s^2}{4\pi}.
\end{equation}
The constant $\alpha_\s$ should be estimated at the scale $\mu_b = m_b = 4.6$\,GeV
where the $\Delta B = 2$ processes are actually measured \cite{UTfit}.
So we have to take into account the renormalization group effect
from the weak scale down to $\mu_b$:
%%% \label{alphasmub} %%%%%%%%%%%%%%%%%%%%%%
\begin{align}
	\label{alphasmub}
	\alpha_\s^{-1}(m_b)
  = \alpha_\s^{-1}(M_Z)
	-\frac{23}{6\pi}\ln\frac{M_Z}{m_b}
	\qquad \lra \qquad
	\alpha_\s(m_b) \approx 0.207
\end{align}
where $\alpha_\s(M_Z) \approx 0.1184$ has been put \cite{2009Bethke}.

Similarly, the each type effective Hamiltonian
for $B_s^0$\,--\,$\bar B_s^0$ mixing %\eqref{BsBsbar} 
are respectively rewritten by replacements $d \leftrightarrow s$ and $31 \leftrightarrow 32$
of \eqref{BBbar2} and \eqref{BBbar2z}.

Combining these results, we obtain the lower bounds for the compactification scale
from the constraint \eqref{BBconstraints}.
First let us assume that only one of the three types of diagrams (LL,  RR,  LR)
gives dominant contribution to the mixing.
Then we get lower bound on the compactification scale by use of the upper bound
on the relevant coefficients $z^q_1$, $\tilde z^q_1$ and $z^q_4$ given
in \eqref{BBconstraints} in the unit of TeV:
\begin{align}
	\begin{array}{rlrl}
	 {\rm LL}~&\!\!\!\!:~~\dis\frac1R
	 \gtrsim \sqrt{\frac{\big|z^d_1\big|}{2.3\times10^{-5}}}~\big[{\rm TeV}\big] &
	 {\rm LL}~&\!\!\!\!:~~\dis\frac1R 
	 \gtrsim \sqrt{\frac{|z^s_1|}{1.1\times10^{-3}}}~\big[{\rm TeV}\big]\\[12pt]
	 {\rm RR}~&\!\!\!\!:~~\dis\frac1R 
	 \gtrsim \sqrt{\frac{\big|\tilde z^d_1\big|}{2.3\times10^{-5}}}~\big[{\rm TeV}\big]
	 \qquad , &\quad
	 {\rm RR}~&\!\!\!\!:~~\dis\frac1R 
	 \gtrsim \sqrt{\frac{|\tilde z^s_1|}{1.1\times10^{-3}}}~\big[{\rm TeV}\big]\\[12pt]
	 {\rm LR}~&\!\!\!\!:~~\dis\frac1R 
	 \gtrsim \sqrt{\frac{\big|z^d_4\big|}{2.1\times10^{-7}}}~\big[{\rm TeV}\big] &
	 {\rm LR}~&\!\!\!\!:~~\dis\frac1R 
	 \gtrsim \sqrt{\frac{|z^s_4|}{1.6\times10^{-5}}}~\big[{\rm TeV}\big]
	\end{array}
\end{align}
Let us note that LR type diagrams yield both of $Q^q_4$ and $Q^q_5$ operators
as is seen in \eqref{BBbar2}.
We can however safely ignore the contribution of $Q^q_5$ to the mixing,
because the coefficients of the operator \eqref{BBbar2zLR5}
are smaller than that of $Q^q_4$
and also because the magnitude of the hadronic matrix element of $Q^q_4$ is known
to be greater than that of $Q^q_5$,
as the constraint for $z^q_4$ is more severe that that for $Q^q_5$ in \eqref{BBconstraints}.
This is why we used the constraint for $z^q_4$ alone to get the lower bound
for the case of LR type diagrams.

Since there is no bulk mass of third generation in this model,
the \lq\lq GIM-like" suppression mechanism from the large bulk masses
which acts much more severe on the contribution of the LR type diagram \cite{2011AKLM} does not occur.
Thus the contribution of the LR-type diagram is not expected 
to be smaller than those of the LL and RR diagram in general.
Actually, for the case of $R_u = \bs1_{3\times3}$ in \eqref{parametrization},
which gives almost the most stringent lower bound from $K^0$\,--\,$\bar K^0$ mixing,
the LR type contribution is dominant for $B_s^0$\,--\,$\bar B_s^0$ mixing
while the LL type contribution is dominant for $B_d^0$\,--\,$\bar B_d^0$ mixing;
\begin{subequations}
\begin{align}
	R^{-1} \gtrsim 1.71\,\big[{\rm TeV}\big]
	\qquad \text{for $B_d^0$\,--\,$\bar B_d^0$ mixing}\ ,\\
	R^{-1} \gtrsim 2.54\,\big[{\rm TeV}\big]
	\qquad \text{for $B_s^0$\,--\,$\bar B_s^0$ mixing}\ .
\end{align}
\end{subequations}

In the second case $R_d = \bs1_{3\times3}$, 
the contributions from the LR and RR type diagram vanish. 
This is because the down-type Yukawa coupling becomes diagonal:
$V_{dL} = V_{dR} = \bs1_{3\times3}$, namely the mixings in the down quark sector disappear. 
Note, however, that the lower bound obtained from the LL type contribution, 
which does not vanish even though $V_{dL} = \bs1_{3\times3}$.
Actually, we obtain the lower bound on $R^{-1}$ for $R_d = \bs1_{3\times3}$;
\begin{subequations}
\begin{align}
	R^{-1} \gtrsim 0.92\,\big[{\rm TeV}\big]
	\qquad \text{for $B_d^0$\,--\,$\bar B_d^0$ mixing}\ ,\\
	R^{-1} \gtrsim 1.79\,\big[{\rm TeV}\big]
	\qquad \text{for $B_s^0$\,--\,$\bar B_s^0$ mixing}\ .
\end{align}
\end{subequations}
This is because $V_{uL}$ relevant for up-type quark mixing also contributes
to the left-handed FCNC current.
Namely, because of the mixing between $Q_{3L}$ and $Q_{6L}$\,($Q_{15L}$),
$U_4$ also contributes to the FCNC vertex \eqref{BBvertexLL}
%and \eqref{BsBsvertexLL}.
Thus even in the case of $V_{dL} = \bs1_{3\times3}$
we get a meaningful lower bound on $M_c$

A comment is given.
The obtained lower bounds are smaller
than what we naively expect assuming that the tree level diagram
relevant for the FCNC process is simply suppressed by $1/M_{\rm c}^2$ \cite{UTfit};
\begin{subequations}
\begin{align}
%	\frac1{M_{\rm c}^2}
% \lesssim \mathcal O\big(10^{-6}\big)\,\big[{\rm TeV}^{-2}\big]
%	\quad \longrightarrow \quad
	M_{\rm c} \gtrsim \mathcal O\big(10^3\big)\,\big[{\rm TeV}\big]
	\qquad \text{for $B_d^0$\,--\,$\bar B_d^0$ mixing}\ ,\\
%%
%	\frac1{M_{\rm c}^2}
% \lesssim \mathcal O\big(10^{-5}\big)\,\big[{\rm TeV}^{-2}\big]
%	\quad \longrightarrow \quad
	M_{\rm c} \gtrsim \mathcal O\big(10^2\big)\,\big[{\rm TeV}\big]
	\qquad \text{for $B_s^0$\,--\,$\bar B_s^0$ mixing}\ .
\end{align}
\end{subequations}
%\begin{align}
%	\frac1{M_{\rm c}^2}
% \lesssim \mathcal O\big(10^{-5}\big)\,\big[{\rm TeV}^{-2}\big]
%	\quad \longrightarrow \quad
%	M_{\rm c} \gtrsim \mathcal O\big(10^2\big)\,\big[{\rm TeV}\big]\ ,
%\end{align}
which is much more stringent than the lower bound we obtained,
in spite of the absence of the suppression by the large bulk masses. 
The obtained lower bounds also are milder
than those from $K^0-\bar K^0$ and $D^0-\bar D^0$ mixings.
This apparent discrepancy may be attributed to the very small mixing
between the third generation and the first two generations.

%%% Summary %%%%%%%%%%%%%%%%%%%%%%%%%%%%%%% 
\section{Summary}
In this paper, we have discussed the flavor mixing
and the resulting FCNC processes in the framework of five dimensional
$SU(3)_{\rm color} \otimes SU(3) \otimes U'(1)$ gauge-Higgs unification scenario. 
As the concrete FCNC processes, 
we have calculated the contributions to $B^0_d-\bar B^0_d$ and $B^0_s-\bar B^0_s$ mixings 
by the non-zero KK gluon exchange at the tree level 
in the light of the recent progress in the measurements of $B^0-\bar B^0$ mixing. 
For the processes with respect to the third generation, 
the \lq\lq GIM-like" suppression mechanism, which is operative for the light first two generation quarks, 
does not work since their bulk mass has to be vanished to realize top quark mass. 
Therefore, we can anticipate large FCNC effects to arise and we are likely to obtain strong constraints for B-physics. 
The prediction of our model is that the lower bounds of compactification scale have been found to be of order 
${\cal O}({\rm TeV})$ which is milder than those obtained from our study of $K^0-\bar{K}^0$ and $D^0-\bar{D}^0$ mixings 
in our previous paper \cite{2010AKLM, 2011AKLM} and from a naive expectation ($\sim 1000~{\rm TeV}$) 
where the dimension six operator is simply suppressed by $1/M_{\rm c}^2$ 
in spite of the absence of the GIM-like suppression by the large bulk masses. 
This is because the smallness of the mixings between 1-3 and 2-3 generations, {\it i.e.} $\theta_2,\theta_3,\theta_2',\theta_3'\ll 1$. 
In our model, they should be small to reproduce the realistic top quark mass $\sim 2M_W$,
and then the induced $\Delta B=2$ effective hamiltonian are strongly suppressed.
Thus the lower bound of compactification scale becomes small.

%%% Acknowledgments %%%%%%%%%%%%%%%%%%%%%%%%%%%
\subsection*{Acknowledgments}
We would like to thank Professor C.S. Lim
for fruitful discussions and valuable suggestions at various stages of this work.
This work was supported in part by the Grant-in-Aid for Scientific Research 
of the Ministry of Education, Science and Culture, No. 21244036 and 
in part by Keio Gijuku Academic Development Funds (N.~M.).

%%%%%%%%%%%%%%%%%%%%%%%%%%%%%%% bibliography %%%%%%%%%%%%%%%%%%%%%%%%%%%%%%%


\begin{thebibliography}{nn}

%%%%
\bibitem{GH} 
  N.~S.~Manton,
  %``A New Six-Dimensional Approach To The Weinberg-Salam Model,''
  Nucl.\ Phys.\ B {\bf 158}, 141 (1979);
%\bibitem{Fairlie}
%\\
  D.~B.~Fairlie,
  %``Higgs' Fields And The Determination Of The Weinberg Angle,''
  Phys.\ Lett.\ B {\bf 82}, 97 (1979), 
  %``Two Consistent Calculations Of The Weinberg Angle,''
  J.\ Phys.\ G {\bf 5}, L55 (1979);
%\\
%\bibitem{Hosotani}  
  Y.~Hosotani,
    %``Dynamical Mass Generation By Compact Extra Dimensions,''
  Phys.\ Lett.\ B {\bf 126}, 309 (1983), 
  %``Dynamical Gauge Symmetry Breaking As The Casimir Effect,''
  Phys.\ Lett.\ B {\bf 129}, 193 (1983), 
  %``DYNAMICS OF NONINTEGRABLE PHASES AND GAUGE SYMMETRY BREAKING,''
  Annals Phys.\  {\bf 190}, 233 (1989).

%%%%%%%%%%%%%%%%%%%%%%%%%%%%%%%%%%%%%%%%%%%%%%%%%%%%%%%%%%%%%%% 
\bibitem{HIL}
  H.~Hatanaka, T.~Inami and C.~S.~Lim,
  %``The gauge hierarchy problem and higher dimensional gauge theories,''
  Mod.\ Phys.\ Lett.\ A {\bf 13}, 2601 (1998). 
%  [arXiv:hep-th/9805067].
%%%%%%%%%%%%%%%%%%%%%%%%%%%%%%%%%%%%%%%%%%%%%%%%%%%%%%%%%%%%%%

\bibitem{ABQ}
  I.~Antoniadis, K.~Benakli and M.~Quiros,
  %``Finite Higgs mass without supersymmetry,''
  New J.\ Phys.\  {\bf 3}, 20 (2001); 
%  [arXiv:hep-th/0108005].
%\\
%\bibitem{GIQ}
  G.~von Gersdorff, N.~Irges and M.~Quiros,
  %``Bulk and brane radiative effects in gauge theories on orbifolds,''
  Nucl.\ Phys.\ B {\bf 635}, 127 (2002); 
%  [arXiv:hep-th/0204223].
%\\
%\bibitem{CNP}
 R.~Contino, Y.~Nomura and A.~Pomarol,
  %``Higgs as a holographic pseudo-Goldstone boson,''
  Nucl.\ Phys.\ B {\bf 671}, 148 (2003); 
%  [arXiv:hep-ph/0306259].
%\\
%\bibitem{LMH}
  C.~S.~Lim, N.~Maru and K.~Hasegawa,
  %``Six dimensional gauge-Higgs unification with an extra space S**2 and the
  %hierarchy problem,''
    J.\ Phys.\ Soc.\ Jap.\  {\bf 77}, 074101 (2008). 
%  arXiv:hep-th/0605180.

\bibitem{HLM}
  K.~Hasegawa, C.~S.~Lim and N.~Maru,
  %``An attempt to solve the hierarchy problem based on gravity gauge Higgs
  %unification scenario,''
  Phys.\ Lett.\ B {\bf 604}, 133 (2004). 
%  [arXiv:hep-ph/0408028].

\bibitem{MY}
  N.~Maru and T.~Yamashita,
  %``Two-loop calculation of Higgs mass in gauge-Higgs unification: 
  %5D  massless QED compactified on S**1,''
  Nucl.\ Phys.\ B {\bf 754}, 127 (2006); 
%  [arXiv:hep-ph/0603237].
%\\
%\bibitem{HMTY}
  Y.~Hosotani, N.~Maru, K.~Takenaga and T.~Yamashita,
  %``Two loop finiteness of Higgs mass and potential in the gauge-Higgs
  %unification,''
  Prog.\ Theor.\ Phys.\  {\bf 118}, 1053 (2007). 
%  [arXiv:0709.2844 [hep-ph]].
  
\bibitem{LM}
  C.~S.~Lim and N.~Maru,
  %``Calculable One-Loop Contributions to S and T Parameters in the
  %Gauge-Higgs Unification,''
  Phys.\ Rev.\  D {\bf 75}, 115011 (2007). 
%  [arXiv:hep-ph/0703017].
  
\bibitem{LHC}
  N.~Maru and N.~Okada,
  %``Gauge-Higgs Unification at LHC,''
  Phys.\ Rev.\  D {\bf 77}, 055010 (2008); 
%  [arXiv:0711.2589 [hep-ph]].
%\\
  N.~Maru,
  %``Finite Gluon Fusion Amplitude in the Gauge-Higgs Unification,''
  Mod.\ Phys.\ Lett.\  A {\bf 23}, 2737 (2008).
%  [arXiv:0803.0380 [hep-ph]].
  
\bibitem{ALM1}
  Y.~Adachi, C.~S.~Lim and N.~Maru,
  %``Finite Anomalous Magnetic Moment in the Gauge-Higgs Unification,''
  Phys.\ Rev.\  D {\bf 76}, 075009 (2007); 
%  [arXiv:0707.1735 [hep-ph]].
%\bibitem{ALM2}
%  Y.~Adachi, C.~S.~Lim and N.~Maru,
  %``More on the Finiteness of Anomalous Magnetic Moment in the Gauge-Higgs
  %Unification,''
    Phys.\ Rev.\  D {\bf 79}, 075018 (2009). 
%  arXiv:0901.2229 [hep-ph] (to appear in PRD).

\bibitem{ALM3}
 Y.~Adachi, C.~S.~Lim and N.~Maru,
  %``Neutron Electric Dipole Moment in the Gauge-Higgs Unification,''
  Phys.\ Rev.\  D {\bf 80}, 055025 (2009). 
%  [arXiv:0905.1022 [hep-ph]].

\bibitem{LMN}
  C.~S.~Lim, N.~Maru and K.~Nishiwaki,
  %``CP Violation due to Compactification,''
  Phys.\ Rev.\  D {\bf 81}, 076006 (2010). 
%  arXiv:0910.2314 [hep-ph].

\bibitem{2010AKLM}
  Y.~Adachi, N,~Kurahashi, C.~S.~Lim and N.~Maru,
  JHEP {\bf 1011} 150 (2010).

\bibitem{2011AKLM}
  Y.~Adachi, N,~Kurahashi, C.~S.~Lim and N.~Maru,
  arXiv:hep-ph/1103.5980.

\bibitem{BN}
  G.~Burdman and Y.~Nomura,
  %``Unification of Higgs and gauge fields in five dimensions,''
  Nucl.\ Phys.\ B {\bf 656}, 3 (2003). 
  
 \bibitem{DPQ}
  A.~Delgado, A.~Pomarol and M.~Quiros,
  %``Electroweak and flavor physics in extensions of the standard model with
  %large extra dimensions,''
  JHEP {\bf 0001}, 030 (2000). 
 % [arXiv:hep-ph/9911252].
    
\bibitem{GW}
  S.~L.~Glashow and S.~Weinberg,
  %``Natural Conservation Laws For Neutral Currents,''
  Phys.\ Rev.\  D {\bf 15}, 1958 (1977).
%  [arXiv:hep-ph/0210257].

%\bibitem{HFAG} 
%E. Barberio {\it et al.} (Heavy Flavor Averaging Group), arXiv:0808.1297. 

\bibitem{2005MSSS}
  G. Martinelli, M. Salvatori, C. A. Scrucca and L. Silvestrini,
  %``Minimal gauge-higgs unification with a flavour symmetry,''
  JHEP {\bf 0510}, 037 (2005).
%  [arXiv:hep-ph/0503179].

\bibitem{2006CCP}
  G.~Cacciapaglia, C.~Csaki and S.~C.~Park,
  %``Fully radiative electroweak symmetry breaking,''
  JHEP {\bf 0603}, 099 (2006). 
%  [arXiv:hep-ph/0510366].

\bibitem{CKMmatrix}
  N. Cabibbo,
  %``Unitary Symmetry and Leptonic Decays,''
  Phys. Rev. Lett. {\bf10}, 531 (1963);
%\\
  M. Kobayashi and T. Maskawa,
  %``CP Violation in the Renormalizable Theory of Weak Interaction,''
  Prog. Theor. Phys. {\bf49}, 652 (1973);
%  [arXiv:hep-ph/9403384].

%\bibitem{Wolfenstein}
%  L. Wolfenstein,
%  %``Parametrization of the Kobayashi-Maskawa Matrix,''
%  Phys. Rev. Lett. {\bf51}, 1945 (1983);
%%\\
%  A. J. Buras, M. E. Lautenbacher and G. Ostermaier,
%  %``Waiting for the top quark mass, $K^+ \to \pi^+\nu\nu^-$,
%  %  $B_s^0$\,--\,$\bar B_s^0$ mixing and CP asymmetries in $B$ decays,''
%  Phys. Rev. {\bf D50}, 3433 (1994);
%%  [arXiv:hep-ph/9403384].
%%\\
%  J. Charles {\it et al.} [ CKMfitter Group Collaboration ],
%  %``CP violation and the CKM matrix: Assessing the impact of the asymmetric $B$ factories,''
%  Eur. Phys. J. {\bf C41}, 1 (2005).
%%  [arXiv:hep-ph/0406184].

\bibitem{ACP}
  K.~Agashe, R.~Contino and A.~Pomarol,
  %``The Minimal composite Higgs model,''
  Nucl.\ Phys.\  B {\bf 719}, 165 (2005). 
%  [arXiv:hep-ph/0412089].
  
\bibitem{UTfit}
  M.~Bona {\it et al.} [ UTfit Collaboration ],
  %``Model-independent constraints on Delta F=2 operators and the scale of new physics,''
  JHEP {\bf 0803}, 049 (2008).
%  [arXiv:0707.0636 [hep-ph]].
  
\bibitem{2011HKMRS}
  K. Huitu, S. Khalil, A. Moursy, S.K. Rai and A. Sabanci,
  arXiv:hep-ph/1105.3087.
  
\bibitem{2009Bethke}
S. Bethke,
%``{\em The 2009 World Average of $\alpha_\s$},''\\
	 Eur. Phys. J. {\bf C64} (2009) 689--703.

      
  
\end{thebibliography}
\end{document}